\documentclass[12pt]{article}
\usepackage{epsf}
\title{ {\bf
The effects of lepton KK modes on the lepton electric dipole
moments in the Randall Sundrum scenario}}
\author{\vspace{1cm}\\
        {\bf E. O. Iltan}
        \thanks{E-mail address:
        eiltan@newton.physics.metu.edu.tr}
 \\
        Physics Department, Middle East Technical University \\
        Ankara, Turkey\\}

\date{}

\begin{document}
\setlength{\baselineskip}{24pt}
\maketitle
\setlength{\baselineskip}{7mm}
\begin{abstract}
We study the charged lepton electric dipole moments in the Randall
Sundrum model where the leptons and the gauge fields are
accessible to the extra dimension. We observe that the electron
(muon; tau) electric dipole moment reaches to the value of the
order of $10^{-26}\, e-cm$ ($10^{-20}\, e-cm$; $10^{-20}\, e-cm$)
with the inclusion of the lepton KK modes.
\end{abstract}
\thispagestyle{empty}
\newpage
\setcounter{page}{1}
\section{Introduction}
The CP violation is among the most interesting physical phenomena
and the electric dipole moments (EDMs) of fermions are important
tools to understand it since EDMs are driven  by the CP violating
interaction. There are various experimental and theoretical works
done in the literature and the experimental results of the
electron, muon and tau EDMs are  $d_e =(1.8\pm 1.2\pm 1.0)\times
10^{-27} e\, cm$ \cite{Commins}, $d_{\mu} =(3.7\pm 3.4)\times
10^{-19} e\, cm$ \cite{Bailey} and $d_{\tau} =(3.1)\times 10^{-16}
e\, cm$ \cite{Groom}, respectively. Furthermore, the experimental
upper bound of neutron EDM has been found as $d_N < 1.1\times
10^{-25} e\, cm$ \cite{Smith1}. From the theoretical point of
view, the source of CP violation in the standard model (SM) is the
complex Cabbibo-Kobayashi-Maskawa (CKM); lepton mixing matrix in
the quark; lepton sector and the calculation of fermion EDMs shows
that their numerical values are negligible in the SM. In work
\cite{Khiplovich}, the quark EDMs have been estimated as $\sim
10^{-30}\, (e-cm)$, which is a small quantity since the non-zero
contribution exists at least in the three loop level. In order to
enhance the fermion EDMs, one needs an alternative source of  CP
violation and the additional contributions coming from the physics
beyond the SM. The multi Higgs doublet models (MHDMs), the
supersymmetric model (SUSY) \cite{Schmidt} are among the possible
models carrying an additional CP phase. The electron EDM has been
predicted of the order of the magnitude of $10^{-32}\, e-cm$ in
the two Higgs doublet model (2HDM), including the tree level
flavor changing neutral currents (FCNC) \cite {Iltmuegam} and, in
this case, the additional CP sources are new complex Yukawa
couplings. The EDMs of fermions have been analyzed in
\cite{IltanExtrEDM} and \cite{IltanSplitEDM}, in the 2HDM with the
inclusion of non-universal extra dimensions and in the framework
of the split fermion scenario. The EDMs of quarks were calculated
in the MHDMs, including the 2HDM in \cite{Weinberg, eril3,Iltant},
the fermion EDMs in the SM, with the inclusion of non-commutative
geometry, have been estimated in \cite{IltanNonCom}, the lepton
EDMs have been studied in the seesaw model in \cite {Bhaskar}, the
EDMs of nuclei, deutron, neutron and some atoms have been
predicted extensively in \cite{Vladimir}, limits on the dipole
moments of leptons have been analyzed in  a left-right symmetric
model and  in $E_6$ superstring models in \cite{Gutierrez}.

In the present work,  we analyze the charged lepton EDMs in the
framework of the 2HDM, with the inclusion of a single extra
dimension, respecting the Randall Sundrum (RS1) scenario
\cite{Rs1, Rs2}. The extra dimensions are introduced to solve the
hierarchy problem between weak and Planck scales. In the RS1
model, the gravity is localized on a 4D boundary, so called hidden
(Planck) brane, and the other fields, including the SM fields,
live on another 4D boundary, so called the visible (TeV) brane.
The warp factor that is an exponential function of the
compactified radius in the extra dimension drives the difference
of induced metrics on these boundaries. With this factor, two
effective scales, the Planck scale $M_{Pl}$ and the weak scale
$m_W$, are connected and the hierarchy problem is solved. If the
SM fields are accessible to the extra dimension, one obtains a
richer phenomenology \cite{Goldberger}-\cite{Moreau2}. The fermion
mass hierarchy can be obtained by considering that the fermions
have different locations in the extra dimension and it is induced
by the Dirac mass term in the Lagrangian \cite{Pamoral2, Grossman,
Huber, Huber2}. \cite{Huber3} is devoted to this hierarchy by
considering that the Higgs field has an exponential profile around
the TeV brane and \cite{Kogan} is devoted to an extensive work on
the bulk fields in various multi-brane models. The different
locations of the fermion fields in the extra dimension can ensure
the flavor violation (FV) and it is carried by the Yukawa
interactions, coming from the SM Higgs-fermion-fermion vertices.
The fermion localization in the RS1 background has been applied to
the high precision measurements of top pair production at the ILC
\cite{Pree} and the various experimental FCNC constraints and the
electro weak precision tests for the location parameters of the
fermions in the extra dimension are analyzed in \cite{Moreau1,
Moreau2}.

Here, we study the charged lepton EDMs in the case that the
leptons and gauge fields are accessible to the extra dimension
with localized leptons in the RS1 background. We observe that the
$d_e$ ($d_\mu$; $d_\tau$) reach to the values of the order of
$10^{-26}\, e-cm$ ($10^{-20}\, e-cm$; $10^{-20}\, e-cm$) with the
inclusion of the KK modes.

The paper is organized as follows: In Section 2, we present EDMs
of charged leptons in the RS1 scenario, in the 2HDM. Section 3 is
devoted to discussion and our conclusions. In Appendix section, we
present how to construct the SM fermions and their KK modes.
\section{Electric dipole moments of charged leptons in the two  Higgs
doublet model, in the RS1 scenario.}
The fermion EDM arises from the CP violating
fermion-fermion-photon effective interaction and, therefore, their
experimental and theoretical search ensure considerable
information about the nature of the CP violation. In the SM, the
CP violation is driven by the complex CKM matrix elements in the
quark sector and a possible lepton mixing matrix in the lepton
sector. The tiny theoretical values of fermion EDMs in the SM
force one to go beyond and the extension of the Higgs sector with
FCNCs at tree level is one of the possibility in order to enhance
their theoretical values. In the present work, we consider the
2HDM, which allows the FCNC at tree level with the complex Yukawa
couplings\footnote{Here, we assume that the CKM type matrix in the
lepton sector does not exist, the charged flavor changing (FC)
interactions vanish and the lepton FV comes from the internal new
neutral Higgs bosons, $h^0$ and $A^0$.}. The additional effect
coming from the extra dimension(s) is another possibility to
enhance the CP violation and, here, we consider the RS1 scenario
with localized charged leptons in the extra dimension. The RS1
background is curved and the corresponding metric reads
\begin{eqnarray}
ds^2=e^{-2\,\sigma}\,\eta_{\mu\nu}\,dx^\mu\,dx^\nu+dy^2\, ,
\label{metric1}
\end{eqnarray}
where $\sigma=k\,|y|$ with the bulk curvature constant $k$ and the
exponential $e^{-k\,|y|}$, with $y=R\,|\theta|$,  is the warp
factor which drives the hierarchy and rescales all mass terms on
the visible brane for $\theta=\pi$. Here $R$ is the
compactification radius in the extra dimension that is
compactified onto $S^1/Z_2$ orbifold with two boundaries, the
hidden (Planck) brane and the visible (TeV) brane. In the RS1
model, all SM fields live in the visible brane, however, the
gravity is accessible to the bulk and it is  considered to be
localized on the hidden brane. With the assumption that the gauge
fields and the fermions are also accessible to the extra
dimension, the particle spectrum is extended and the physics
becomes richer. In the present work, we consider this scenario
with the addition of $Z_2$ invariant\footnote{The fermions have
two possible transformation properties under the orbifold $Z_2$
symmetry, $Z_2\psi=\pm\gamma_5 \psi$ and the combination
$\bar{\psi}\psi$ is odd. The $Z_2$ invariant mass term is obtained
if the $Z_2$ odd scalar field is coupled to the combination
$\bar{\psi}\psi$.} Dirac mass term to the lagrangian of bulk
fermions which results in the fermion localization in the extra
dimension \cite{Hisano, Hewett, Pamoral2, Huber4, Grossman,
Huber2,Huber3}
\begin{eqnarray}
{\cal{S}}_m=-\int d^4x \int dy \,\sqrt{-g}\,m(y)\,\bar{\psi}\psi
\label{massterm} \, ,
\end{eqnarray}
where $m(y)=m\frac{\sigma'(y)}{k}$ with
$\sigma'(y)=\frac{d\sigma}{dy}$ and
$g=Det[g_{MN}]=e^{-8\,\sigma}$, $M,N=0,1,... ,4$. Here the bulk
fermion is expanded as
\begin{eqnarray}
\psi(x^\mu,y)=\frac{1}{\sqrt{2\,\pi\,R}}\,\sum_{n=0}^\infty\,
\psi^{(n)}(x^\mu)\, e^{2\,\sigma}\, \chi_n(y) \label{psiKK} \, ,
\end{eqnarray}
with the normalization
\begin{eqnarray}
\frac{1}{2\,\pi\,R}\,\int_{-\pi\,R}^{\pi\,R}\,dy\,e^\sigma \,
\chi_n(y)\,\chi_m(y)=\delta_{nm} \label{norm} \, .
\end{eqnarray}
Using the Dirac equation and the normalization condition, the zero
mode fermion is obtained as follows:
\begin{eqnarray}
\chi_0(y)=N_0\, e^{-r\,\sigma}\label{0mode} \, ,
\end{eqnarray}
where $r=m/k$ and the normalization constant $N_0$ reads
\begin{eqnarray}
N_0=\sqrt{\frac{k\,\pi \,R\,(1-2\,r)}{e^{k\,\pi
\,R\,(1-2\,r)}-1}}\label{norm0mode} \, .
\end{eqnarray}
On the other hand
\begin{eqnarray}
\chi'_0(y)=e^{-\frac{\sigma}{2}}\,\chi_0(y) \,, \label{0modep}
\end{eqnarray}
is the appropriately normalized solution and it is localized in
the extra dimension.  The parameter $r$ is responsible for the
localization and for $r>\frac{1}{2}$ ($r<\frac{1}{2}$) it is
localized near the hidden (visible) brane.

The action which is responsible for the charged lepton EDMs in the
RS1 background is:
\begin{eqnarray}
{\cal{S}}_{Y}= \int d^5x \sqrt{-g} \,\Bigg( \xi^{E}_{5\,
ij}\,\bar{l}_{i L} \phi_{2} E_{j R} + h.c. \Bigg)\, \delta(y-\pi
R) \,\,\, , \label{yukawalagrangian}
\end{eqnarray}
where $L$ and $R$ denote chiral projections $L(R)=1/2(1\mp
\gamma_5)$, $\phi_{2}$ is the new scalar doublet, $l_{i L}$ ($E_{j
R}$) are lepton doublets (singlets), $\xi^{E}_{5\,ij}$, with
family indices $i,j$ , are the complex Yukawa couplings in five
dimensions and they induce the FV interactions in the lepton
sector. Here, we assume that the Higgs doublet $\phi_1$, living on
the visible brane, has non-zero vacuum expectation value to ensure
the ordinary masses of the gauge fields and the fermions, however,
the second doublet, that lies also on the visible brane, has no
vacuum expectation value:
\begin{eqnarray}
\phi_{1}=\frac{1}{\sqrt{2}}\left[\left(\begin{array}{c c}
0\\v+H^{0}\end{array}\right)\; + \left(\begin{array}{c c} \sqrt{2}
\chi^{+}\\ i \chi^{0}\end{array}\right) \right]\, ;
\phi_{2}=\frac{1}{\sqrt{2}}\left(\begin{array}{c c} \sqrt{2}
H^{+}\\ H_1+i H_2 \end{array}\right) \,\, , \label{choice}
\end{eqnarray}
and
\begin{eqnarray}
<\phi_{1}>=\frac{1}{\sqrt{2}}\left(\begin{array}{c c}
0\\v\end{array}\right) \,  \, ; <\phi_{2}>=0 \,\, .
\label{choice2}
\end{eqnarray}
The gauge and $CP$ invariant Higgs potential which spontaneously
breaks  $SU(2)\times U(1)$ down to $U(1)$ reads:
\begin{eqnarray}
V(\phi_1, \phi_2 )&=&c_1 (\phi_1^+ \phi_1-v^2/2)^2+ c_2 (\phi_2^+
\phi_2)^2 \nonumber \\ &+&  c_3 [(\phi_1^+ \phi_1-v^2/2)+ \phi_2^+
\phi_2]^2 + c_4 [(\phi_1^+ \phi_1)
(\phi_2^+ \phi_2)-(\phi_1^+ \phi_2)(\phi_2^+ \phi_1)] \nonumber \\
&+& c_5 [Re(\phi_1^+ \phi_2)]^2 + c_{6} [Im(\phi_1^+ \phi_2)]^2
+c_{7} \, , \label{potential}
\end{eqnarray}
with constants $c_i, \, i=1,...,7$. The choice eq.(\ref{choice2})
and the potential eq.(\ref{potential}) leads to the fact that the
SM particles are collected in the first doublet and the new
particles in the second one. This is the case that no mixing
occurs between CP-even neutral Higgs bosons $H^0$ and $h^0$ in the
tree level and $H_1$ and $H_2$ in  eq. (\ref{choice}) are obtained
as the mass eigenstates $h^0$ and $A^0$, respectively.

The lepton doublet and singlet fields in eq.
(\ref{yukawalagrangian}) is expanded to their KK modes as
%
\begin{eqnarray}
l_{i L}(x^\mu,y)&=&\frac{1}{\sqrt{2\,\pi\,R}}\, e^{2\,\sigma}\,
l_{i L}^{(0)}(x^\mu)\, \chi_{i\,L 0}(y)\nonumber
\\ &+& \frac{1}{\sqrt{2\,\pi\,R}}\,\sum_{n=1}^\infty\, e^{2\,\sigma}\,
\Bigg( l_{i L}^{(n)}(x^\mu)\, \chi^{l}_{i\,L n}(y)+l_{i
R}^{(n)}(x^\mu)\, \chi^{l}_{i\,R n}(y)\Bigg)\, , \nonumber
\\
E_{j R}(x^\mu,y)&=& \frac{1}{\sqrt{2\,\pi\,R}}\,
e^{2\,\sigma}\,E_{j R}^{(0)}(x^\mu)\, \chi_{j\,R 0}(y)\nonumber
\\ &+&\frac{1}{\sqrt{2\,\pi\,R}}\,\sum_{n=0}^\infty\,
e^{2\,\sigma}\,\Bigg( E_{j R}^{(n)}(x^\mu)\, \chi^E_{j\,R n}(y)+
E_{j L}^{(n)}(x^\mu)\, \chi^E_{j\,L n}(y)\Bigg) \label{leptonKK}
\, ,
\end{eqnarray}
where the zero mode leptons $\chi_{i\,L 0}(y)$ and  $\chi_{j\,R
0}(y)$ are given in eq. (\ref{0mode}) with the replacements
$r\rightarrow r_{iL}$ and $r\rightarrow r_{iR}$, respectively. For
the effective Yukawa coupling $\xi^{E}_{ij}$, we  integrate out
the Yukawa interaction eq.(\ref{yukawalagrangian}) over the fifth
dimension by taking the zero mode lepton doublets, singlets, and
neutral Higgs fields $S=h^0,A^0$:
\begin{eqnarray}
V^0_{RL\,ij}&=&\frac{\xi^{E}_{5\,
ij}}{2\,\pi\,R}\,\int_{-\pi\,R}^{\pi\,R}\,dy\,
\chi_{iR0}(y)\,\chi_{jL0}(y)\,\delta(y-\pi R) \nonumber \\ &=&
\frac{e^{-k\,\pi\,R\,(r_{iR}+r_{jL})}\,k\,
\sqrt{(1-2\,r_{iR})\,(1-2\,r_{jL})}}
{\sqrt{(e^{k\,\pi\,R\,(1-2\,r_{iR})}-1)\,
(e^{k\,\pi\,R\,(1-2\,r_{jL})}-1)}}\, \xi^{E}_{5\,
ij}\label{yukawa0LR} \, .
\end{eqnarray}
Here, we embed the vertex factor $V^0_{RL\,(LR)\,ij}$ into the
coupling $\xi^{E}_{ij}\,\Big((\xi^{E}_{ij})^\dagger\Big)$ and fix
the numerical value of
$\xi^{E}_{ij}\,\Big((\xi^{E}_{ij})^\dagger\Big)$ by assuming that
the coupling $\xi^{E}_{5\, ij}$ in five dimension is flavor
dependent. In this case the hierarchy of new Yukawa couplings is
not related to the lepton field locations. On the other hand, we
need the vertex factors due to $S$-KK mode charged lepton-charged
lepton couplings $V^n_{LR\,(RL)\,ij}$ since charged lepton KK
modes exist in the internal line (see fig.1). After the
integration of the Yukawa interaction eq.(\ref{yukawalagrangian})
over the fifth dimension, the $S$-zero mode lepton singlet
(doublet)-KK mode lepton doublet (singlet) vertex factor (see
appendix for the construction of KK mode charged lepton doublets
and singlets) reads
\begin{eqnarray}
V^n_{RL\,(LR)\,ij}=\frac{\xi^{E}_{5\, ij}\,\Big((\xi^{E}_{5\,
ij})^{\dagger}\Big) }{2\,\pi\,R}\,\int_{-\pi\,R}^{\pi\,R}\,dy\,\,
\chi_{iR0\,(iL0)}(y)\,\, \chi_{jLn\,(jRn)}(y)\,\,\delta(y-\pi\,
R)\label{intyukawa0KK} \, ,
\end{eqnarray}
and
\begin{eqnarray}
V^n_{RL\,\,ij}&=&\frac{ N_{Ln}\, e^{k \pi R\,(1/2-r_{iR})}\,
\Bigg( J_{\frac{1}{2}-r_{jL}} (e^{k \pi R}\,x_{nL})+c_L\,
Y_{\frac{1}{2}-r_{jL}}(e^{k \pi R}\,x_{nL})\Bigg) } {\pi R \sqrt{
\frac{e^{k\,\pi \,R\,(1-2\,r_{iR})}-1}{k\,\pi
\,R\,(1-2\,r_{iR})}}}\,\xi^{E}_{5\, ij}
\nonumber \, , \\
V^n_{LR\,\,ij}&=& \frac{N_{Rn}\, e^{k \pi R\,(1/2-r_{iL})}\,
\Bigg( J_{\frac{1}{2}+r_{jR}} (e^{k \pi R}\,x_{nR})+c_R\,
Y_{\frac{1}{2}+r_{jR}} (e^{k \pi R}\,x_{nR})\Bigg)}{\pi R
\sqrt{\frac{e^{k\,\pi \,R\,(1-2\,r_{iL})}-1}{k\,\pi
\,R\,(1-2\,r_{iL})}}}\, (\xi^{E}_{5\,
ij})^{\dagger}\label{yukawa0LKKR} \, ,
\end{eqnarray}
where $c_L$ ($c_R$) is given in eq.(\ref{cL}) (eq.(\ref{cR})) with
the replacements $r\rightarrow r_{jL}$ ($r\rightarrow r_{jR}$).
Here the effective $S$-zero mode lepton singlet (doublet)-KK mode
lepton doublet (singlet) coupling $\xi^{E\, n}_{ij}\,(\xi^{E\,
n}_{ij})^\dagger$ reads
\begin{eqnarray}
\xi^{E\, n}_{ij}\,\Big((\xi^{E\,n}_{ij})^\dagger\Big)=
\frac{V^n_{RL\,(LR)\,ij}}{V^0_{RL\,(LR)\,ij}}\, \xi^{E}_{ij} \,\,
. \label{set2Yukawan}
\end{eqnarray}
Notice that the strengths of $S$-KK mode charged lepton-charged
lepton couplings are regulated by the locations of the lepton
fields and we use two different sets, Set I and Set II\footnote{In
this scenario, the source of FV is not related to the different
locations of the fermion fields in the extra dimension (see
\cite{KAgashe, EBlechman}) but it is carried by the new Yukawa
couplings in four dimensions and the additional effect due to the
extra dimension is the enhancement in the physical quantities of
the processes studied.}.

%
\begin{table}[h]
        \begin{center}
\begin{tabular}{|c|c|c|c|c|}
  \hline
    & SET I  & SET  II \\
  \hline \hline
& \,$r_L$ \,\,\,\,\,  $r_R$  & \,$r_L$  \,\,\,\,\,$r_R$  \\
\hline\hline
  e & -0.4900 \,\, 0.8800 & -1.0000 \,\, 0.8860 \\ \hline
  $\mu$  & -0.4900 \,\, 0.7160 & -1.0000 \,\, 0.7230 \\ \hline
  $\tau$ & -0.4900 \,\,0.6249 & -1.0000 \,\,  0.6316 \\ \hline
  \hline
\end{tabular}
\end{center}
\caption{Two possible locations of charged lepton fields. Here
$r_L$ and $r_R$ are left handed and right handed lepton field
location parameters, respectively.} \label{set}
\end{table}
%

\begin{figure}[htb]
\vskip -5.3truein \centering\,\,\,\,\,\,\,\,\,\,\,\,\,\,\,
\epsfxsize=5.0in \leavevmode\epsffile{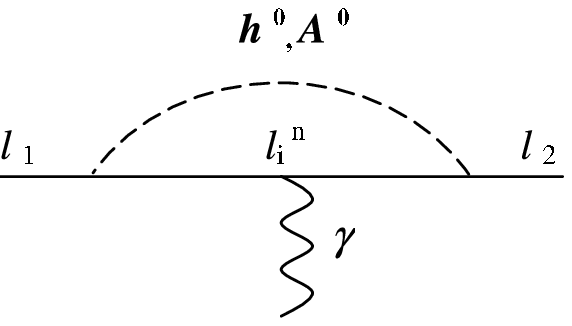} \vskip 10truein
\caption[]{One loop diagrams contributing to EDM of $l$-lepton due
to the neutral Higgs bosons $h^0$ and $A^0$ in the 2HDM. Wavy
(dashed-solid) line represents the electromagnetic field ($h^0$ or
$A^0$ fields-charged lepton fields and their KK modes).}
\label{fig1}
\end{figure}

The effective EDM interaction for a fermion $f$ is reads
\begin{eqnarray}
{\cal L}_{EDM}=i d_f \,\bar{f}\,\gamma_5 \,\sigma^{\mu\nu}\,f\,
F_{\mu\nu} \,\, , \label{EDM1}
\end{eqnarray}
where $F_{\mu\nu}$ is the electromagnetic field tensor, '$d_{f}$'
is EDM of the fermion and it is a real number by hermiticity (see
Fig. \ref{fig1} for the 1-loop diagrams which contribute to the
EDMs of fermions).

Now, we present the charged lepton EDMs with the addition of KK
modes in the framework of the RS1 scenario. Since there is no CKM
type lepton mixing matrix according to our assumption, only the
neutral Higgs part gives a contribution to their EDMs and
$l$-lepton EDM '$d_l$' $(l=e,\,\mu,\,\tau)$  can be calculated as
a sum of contributions coming from neutral Higgs bosons $h_0$ and
$A_0$. For $l=e,\mu$ and we get\footnote{In the following we use
the dimensionful coupling $\bar{\xi}^{E}_{N,ij}$ in four
dimensions, with the definition $\xi^{E}_{N,ij}=\sqrt{\frac{4\,
G_F}{\sqrt{2}}}\, \bar{\xi}^{E}_{N,ij}$ where N denotes the word
"neutral".},
\begin{eqnarray}
d_l&=& -\frac{i G_F}{\sqrt{2}} \frac{e}{32\pi^2}\, Q_{\tau}\Bigg\{
\frac{1}{m_{\tau}}\, ((\bar{\xi}^{E\,*}_{N,l\tau})^2-
(\bar{\xi}^{E}_{N,\tau l})^2)\, \Bigg ( (F_1 (y_{h^0})-F_1
(y_{A_0})) \Bigg) \nonumber \\
&+& \sum_{n=1}^{\infty}\,((\bar{\xi}^{E\,n\,*}_{N,l\tau})^2-
(\bar{\xi}^{E\,n}_{N,\tau l})^2)\, \Bigg ( G(y_{nL, h^0}, y_{nR,
h^0})-G(y_{nL, A^0}, y_{nR, A^0})\Bigg )\Bigg\}, \label{emuEDM}
\end{eqnarray}
where
\begin{eqnarray}
G(y_{nL, S}, y_{nR, S})&=&G_1(y_{nL, S}, y_{nR, S})+G_1(y_{nR, S},
y_{nL, S})+G_2(y_{nL, S}, y_{nR, S}) \nonumber \\ &+& G_2(y_{nR,
S}, y_{nL, S})\,, \label{emuEDMfunct1}
\end{eqnarray}
with
\begin{eqnarray}
F_1 (w)&=&\frac{w\,(3-4\,w+w^2+2\,ln\,w)}{(-1+w)^3}\nonumber \,\, , \\
\label{functions1}
\end{eqnarray}
and
\begin{eqnarray}
G_1(y_{nL, S}, y_{nR, S})&=&\frac{y_{nR, S}\,\Bigg(
m_{nL}\,(y_{nR, S}-1)\,y_{nR, S}- m_{nR}\,\Big(y_{nL, S}\,(y_{nR,
S}-2)+y_{nR, S} \Big) \Bigg)}{m_S^2\,(y_{nR,
S}-y_{nL, S})^2\,(1-y_{nR, S})^2 }\,ln\,y_{nR, S}\,,\nonumber \\
G_1(y_{nR, S}, y_{nL, S})&=&G_1(y_{nL, S}, y_{nR, S})|_{y_{nL,
S}\Leftrightarrow y_{nR, S};\, m_{nL}\Leftrightarrow m_{nR}}\, ,
\nonumber \\
G_2(y_{nL, S}, y_{nR, S})&=&\frac{y_{nL,
S}\,m_{nL}}{m_S^2\,(y_{nR, S}-y_{nL, S})\,(1-y_{nL, S})}\, ,
\nonumber \\
G_2(y_{nR, S}, y_{nL, S})&=&G_2(y_{nL, S}, y_{nR, S})|_{y_{nL,
S}\Leftrightarrow y_{nR, S};\, m_{nL}\Leftrightarrow m_{nR}}\, .
\label{emuEDMfunct2}
\end{eqnarray}
The tau lepton EDM reads
\begin{eqnarray}
d_{\tau}&=& -\frac{i G_F}{\sqrt{2}} \frac{e}{16\pi^2}\,
Q_{\tau}\,\Bigg\{ \frac{1}{m_{\tau}}\,
((\bar{\xi}^{E\,*}_{N,\tau\tau})^2- (\bar{\xi}^{E}_{N,\tau
\tau})^2)\, \int_0^1\, dx\, \int_0^{1-x}\, dy\, \Bigg(
(x-1)\,\Big(
\frac{y_{h^0}}{L_{h^0}}-\frac{y_{A^0}}{L_{A^0}}\Big)\nonumber \\
&+& \sum_{n=1}^{\infty}\,  \int_0^1\, dx\, \int_0^{1-x}\, dy\,
\Bigg( (\bar{\xi}^{E\,n}_{N,\tau \tau})^2
\Big(m_{nR}\,y+m_{nL}\,(1-x-y)\Big) \Big(
\frac{1}{m^2_{h^0}\,L_{n,h^0}}-\frac{1}{m^2_{A^0}\,L_{n,A^0}}\Big)
\nonumber\\ &-& (\bar{\xi}^{E\,n\,*}_{N,\tau \tau})^2
\Big(m_{nL}\,y+m_{nR}\,(1-x-y)\Big) \Big(
\frac{1}{m^2_{h^0}\,L'_{n,h^0}}-\frac{1}{m^2_{A^0}\,L'_{n,A^0}}\Big)
\Bigg) \Bigg\} \, ,\label{tauEDM}
\end{eqnarray}
where
\begin{eqnarray}
L_{S}&=&x+(x-1)^2\,y_{S}\, ,
\nonumber \\
L_{n,S}&=&x+x\,(x-1)\,y_{S}+y\, y_{nR, S}+(1-x-y)\,y_{nL, S}\, ,
\nonumber \\
L'_{n,S}&=&L_{n,S}|_{{y_{nL, S}\Leftrightarrow y_{nR, S}} }  \, .
\label{tauEDMfunct1}
\end{eqnarray}
Here $y_S=\frac{m^2_\tau}{m^2_S}$, $y_{nL\,(nR),
S}=\frac{m^2_{nL\,(nR)}}{m^2_S}$ and $Q_{\tau}$ is the tau lepton
charge. In eqs. (\ref{emuEDM}) and (\ref{tauEDM}) we take into
account only the internal $\tau$-lepton contribution, respecting
our assumption that the Yukawa couplings $\bar{\xi}^{E}_{N, ij},\,
i,j=e,\mu$, are small compared to $\bar{\xi}^{E}_{N,\tau\, i}\,
i=e,\mu,\tau$ (see Discussion section for details).  Notice that,
we make our calculations in arbitrary photon four momentum square
$q^2$ and take $q^2=0$ at the end. For the Yukawa couplings we
used the parametrization
\begin{eqnarray}
\bar{\xi}^{E\,n}_{N,\tau l}=|\bar{\xi}^{E\,n}_{N,\tau l}|\, e^{i
\theta_{l}} \,\, , \label{xi2}
\end{eqnarray}
where $l=e,\mu, \tau$. Here, $\theta_{l }$ is the CP violating
parameter which is the source of the lepton EDM\footnote{The
Yukawa factors in  eq.(\ref{emuEDM}) can be written as
\begin{eqnarray}
((\bar{\xi}^{E\,n\,*}_{N,l\tau})^2-(\bar{\xi}^{E\,n\,}_{N,\tau
l})^2)=-2\,i sin\,2\theta_{l}\, |\bar{\xi}^{E\,n\,}_{N,\tau
l}|^2\,.
\end{eqnarray}
}.
\section{Discussion}
In this section, we study the effects lepton KK modes on the EDMs
of charged leptons in the framework of the RS1 scenario, with
extended Higgs sector. The source of fermion EDMs is the CP
violating interaction that is arisen from a CP violating phase.
Here, we assume that  this phase comes from the complex Yukawa
couplings appearing in the tree level fermion-fermion-new Higgs
interactions, in the framework of the 2HDM. In the case of charged
leptons, the leptonic complex Yukawa couplings
$\bar{\xi}^E_{N,ij}, i,j=e, \mu, \tau$ are responsible for the
EDMs and they are free parameters in the model considered. Here,
we expect that the Yukawa couplings $\bar{\xi}^{E}_{N,ij},\,
i,j=e,\mu $ are weak compared to $\bar{\xi}^{E}_{N,\tau\, i}\,
i=e,\mu,\tau$ and the couplings $\bar{\xi}^{E}_{N,ij}$ in four
dimensions are symmetric with respect to the indices $i$ and $j$.
Finally, we consider that the coupling $\bar{\xi}^{E}_{N,\tau\,
e}$, ($\bar{\xi}^{E}_{N,\tau\, \mu}$,
$\bar{\xi}^{E}_{N,\tau\,\tau}$) is dominant among the others. This
is the case that the tau lepton and its KK mode appear in the
internal line (see Fig.1). The numerical value of the coupling
$\bar{\xi}^{E}_{N,\tau \mu}$ is chosen  by respecting the
experimental uncertainty, $10^{-9}$, in the measurement of the
muon anomalous magnetic moment \cite{BNL} (see \cite{Iltananomuon}
for details)\footnote{In \cite{Iltananomuon}, the upper limit of
the coupling $\bar{\xi}^{E}_{N,\tau \mu}$ is estimated as $\sim
30\, (GeV)$ in the framework of the 2HDM and here, we take the
numerical value which is less than this quantity.}. This upper
limit and the experimental upper bound of BR of $\mu\rightarrow e
\gamma$ decay, BR $\leq 1.2\times 10^{-11}$, can give clues about
the numerical value of the coupling $\bar{\xi}^{E}_{N,\tau e}$
(see \cite{Iltmuegam}) and we take it of the order of $10^{-2}\,
(GeV)$. For the coupling $\bar{\xi}^{E}_{N,\tau \tau}$ there is no
stringent prediction and we consider an intermediate value which
is greater than the coupling $\bar{\xi}^{E}_{N,\tau \mu}$. For the
CP violating parameter which drives the EDM interaction we choose
the range, $0.1 \geq \sin\,\theta_{e\, (\mu,\, \tau)} \geq 0.7$.

In the present work, we study the charged lepton EDMs in the RS1
background with the assumption that the leptons are also
accessible to the extra dimension. The inclusion of extra
dimensions brings additional contributions which come from the KK
modes of leptons in the 4D effective theory after the
compactification. Here, we consider that the lepton fields are
localized in the extra dimension with the help of the Dirac mass
term $m_l=r \sigma'$, $\sigma=k\,|y|$ (eq.(\ref{massterm})) in the
action. This is the case that the SM leptons, the right and left
handed parts are located in the extra dimension with exponential
profiles (see eq.(\ref{0mode})) which makes it possible to explain
the different flavor mass hierarchy (see Appendix section for the
construction of the SM fields and their masses).

The gauge sector of the model should live in the extra dimension
necessarily if the leptons exist in the bulk and, therefore, their
KK modes appear after the compactification of the extra dimension.
These KK modes result in additional FCNC effects at tree level
coming from the couplings with charged leptons and they should be
suppressed even for low KK masses, by choosing the lepton location
parameters $c_L$ ($c_R$) appropriately (see the discussion given
in \cite{Moreau1, Moreau2}). Here, we use two different sets of
locations of charged leptons (Table \ref{set}) in order to obtain
the masses  of different flavors and we verify the various
experimental FCNC constraints with KK neutral gauge boson masses
as low as few TeVs.
In both sets, we estimate the right handed locations of leptons by
choosing the left handed charged lepton locations as the same. For
the case that the left handed charged lepton locations near to the
visible brane (see set II), the strengths of the couplings of
leptons with the new Higgs doublet living in the 4D brane becomes
stronger and, therefore, the physical quantities related to these
couplings enhance.

Throughout our calculations we use the input values given in Table
(\ref{input}).
\begin{table}[h]
        \begin{center}
        \begin{tabular}{|l|l|}
        \hline
        \multicolumn{1}{|c|}{Parameter} &
                \multicolumn{1}{|c|}{Value}     \\
        \hline \hline
        $m_{\mu}$                   & $0.106$ (GeV) \\
        $m_{\tau}$                  & $1.78$ (GeV) \\
        $m_{h^0}$           & $100$   (GeV)  \\
        $m_{A^0}$           & $200$   (GeV)  \\
        $G_F$             & $1.16637 10^{-5} (GeV^{-2})$  \\
        \hline
        \end{tabular}
        \end{center}
\caption{The values of the input parameters used in the numerical
          calculations.}
\label{input}
\end{table}
Furthermore, the curvature parameter $k$ and the compactification
radius $R$ are the additional free parameters of the theory. Here
we take $k\,R=10.83$ and consider in the region $10^{17}\leq k
\leq 10^{18}$ (see the discussion in appendix and the work
\cite{Huber2}).

In Fig.\ref{EDMek}, we present the parameter $k$ dependence of the
electron EDM $d_e$, for $\bar{\xi}^{E}_{N,\tau e}=0.01\,(GeV)$ and
two different values of the CP violating parameter $sin\,\theta$.
Here, the lower-upper solid (dashed, small dashed) line represents
the $d_e$ for $sin\,\theta=0.1-0.5$ without KK modes (with KK
modes set I, II). It is observed that the $d_e$ is of the order of
$10^{-29}-10^{-28}\, e-cm$ for $sin\,\theta=0.1-0.5$ without KK
modes. The inclusion of the KK modes enhances the $d_e$ 50
times-two orders for the set I-II for the values of the curvature
scale $k\sim 10^{17}\, (GeV)$ and this enhancement becomes weak
for $k\sim 10^{18}\, (GeV)$. For the set II, the enhancement in
the $d_e$ is almost two times larger compared to the set I case.
This is due to the fact that the left handed leptons (zero and KK
modes) are near to the visible brane and their couplings to the
new Higgs scalars become stronger for set II case. The
experimental upper limit is $d_e =(1.8\pm 1.2\pm 1.0)\times
10^{-27} e\, cm$ and this numerical value can be reached even for
small values of the CP parameter $sin\,\theta \sim 0.1$ with the
inclusion of charged lepton KK modes.

Fig. \ref{EDMmuk} is devoted to the parameter $k$ dependence of
the $d_\mu$, for $\bar{\xi}^{E}_{N,\tau \mu}=1\,(GeV)$ and two
different values of the CP violating parameter $sin\,\theta$.
Here, the lower-upper solid (dashed, small dashed) line represents
the $d_\mu$ for $sin\,\theta=0.1-0.5$ without KK modes (with KK
modes set I, II). We observe that the $d_\mu$ is of the order of
$10^{-25}-10^{-24}\, e-cm$ for $sin\,\theta=0.1-0.5$ without KK
modes. With the inclusion of the KK modes the $d_\mu$ increases to
the values of the order of $10^{-23}; 2.0\times 10^{-23}\, e-cm$
for $sin\,\theta=0.1$ and $5.0\times 10^{-23}; 10^{-22}\, e-cm$
for $sin\,\theta=0.5$ in the case of  the set I; set II, for the
values of the curvature scale $k\sim 10^{17}\, (GeV)$. With the
choice of $\bar{\xi}^{E}_{N,\tau \mu}=10\,(GeV)$, which is the
numerical value near to the upper limit that is obtained by
respecting the experimental uncertainty, $10^{-9}$, in the
measurement of the muon anomalous magnetic moment (see
\cite{Iltananomuon}), the $d_{\mu}$ is reached to the value
$10^{-20}\,e-cm$ for $sin\,\theta \sim 0.5$, with the inclusion of
charged lepton KK modes in the case of set II and $k\sim 10^{17}\,
(GeV)$. This is a numerical value near to the experimental upper
limit $d_{\mu} =(3.7\pm 3.4)\times 10^{-19}\, e-cm$.

Fig.\ref{EDMtauk} represents the parameter $k$ dependence of the
$d_\tau$, for $\bar{\xi}^{E}_{N,\tau \tau}=10\,(GeV)$ and two
different values of the CP violating parameter $sin\,\theta$.
Here, the lower-upper solid (dashed, small dashed) line represents
the $d_\tau$ for $sin\,\theta=0.1-0.5$ without KK modes (with KK
modes set I, II). Here it is observed that the $d_\tau$ is of the
order of $10^{-23}-10^{-22}\, e-cm$ for $sin\,\theta=0.1-0.5$
without KK modes. The inclusion of the KK modes causes that
$d_\tau$ is enhanced to the values of the order of $10^{-21};
2.0\times 10^{-21}\, e-cm$ for $sin\,\theta=0.1$ and $5.0\times
10^{-21}; 10^{-20}\, e-cm$ for $sin\,\theta=0.5$ in the case of
the set I; set II, for the values of the curvature scale $k\sim
10^{17}\, (GeV)$. The experimental upper limit of $d_{\tau}$,
$|d_{\tau}| <(3.1)\times 10^{-16}\, e-cm$, is almost two orders
far from the theoretical value obtained, even with the strong
coupling $\bar{\xi}^{E}_{N,\tau \tau}\sim 100\,(GeV)$ and it needs
more sensitive experimental measurements.

Now, we analyze the CP violating parameter $sin\,\theta$
dependence of the charged lepton EDMs, for completeness.

In Fig.\ref{EDMemutausintet}, we present the parameter
$sin\,\theta$ dependence of the $d_e$; $d_\mu$; $d_\tau$, for
$\bar{\xi}^{E}_{N,\tau e}=0.01\,(GeV)$; $\bar{\xi}^{E}_{N,\tau
\mu}=1.0\,(GeV)$; $\bar{\xi}^{E}_{N,\tau \tau}=10\,(GeV)$ and for
$k=10^{18}\, (GeV)$. Here, the lower-intermediate-upper solid
(dashed, small dashed) line represents the $d_e$-$d_\mu$-$d_\tau$
without KK modes (with KK modes set I, II). Here it is observed
that the EDMs are not so much sensitive to the location of lepton
fields in the bulk for the large values of the curvature parameter
$k$, $k\sim 10^{18}\, (GeV)$ and the enhancements in the EDMs are
$\%6 \,(\%12)$ for set I (set II). This sensitivity becomes weak
for the small value of the CP violating parameter $sin\,\theta$.

With the more accurate experimental investigation of the charged
lepton EDMs, it will be possible to understand the mechanism
behind the CP violation and one will get powerful information
about the effects of warped extra dimensions, if they exist.
\section{Appendix}
The SM fermions are constructed by considering the $SU(2)_L$
doublet $\psi_L$ and the singlet $\psi_R$, satisfying two separate
$Z_2$ projection conditions: $Z_2\psi_R=\gamma_5 \psi_R$ and
$Z_2\psi_L=-\gamma_5 \psi_L$ (see for example \cite{Hisano}). The
zero mode fermions can get mass through the $Z_2$ invariant left
handed fermion-right handed fermion-Higgs interaction,
$\bar{\psi}_R\,\psi_L\, H $\footnote{Here, we consider different
location parameters $r$ for each left handed and right handed part
of different flavors.} and, one gets the location parameters of
the left and the right handed parts of fermions in order to obtain
the current masses of fermions of different flavors. If we
consider that the SM Higgs field lives on the visible brane, the
masses of fermions are calculated by using the integral
\begin{eqnarray}
m_i=\frac{1}{2\,\pi\,R}\,\int_{-\pi\,R}^{\pi\,R}\,dy\,\lambda_5\,
\chi_{iL0}(y)\,\chi_{iR0}(y)\,<H>\,\delta(y-\pi\, R) \label{mi} \,
,
\end{eqnarray}
where $\lambda_5$ is the coupling in five dimensions and it can be
parametrized  in terms of the one in four dimensions, the
dimensionless coupling $\lambda$, $\lambda_5=\lambda/\sqrt{k}$.
Here the expectation value of the Higgs field $<H>$ reads
$<H>=v/\sqrt{k}$ where $v$ is the vacuum expectation value,
$v=0.043\,M_{Pl}$, in order to provide the measured gauge boson
masses \cite{Huber2} and choose $k\,R=10.83$ in order to get the
correct effective scale on the visible brane, i.e.,
$M_W=e^{-\pi\,k\,R}\, M_{pl}$ is of the order of TeV.

Since the EDMs of fermions exist at least in the one loop level,
there appears the $S$-charged lepton-KK charged lepton vertices.
The $Z_2$ the projection condition $Z_2\psi=-\gamma_5 \psi$, used
to construct the left handed fields on the branes and as a result,
the left handed zero and KK modes appear, the right handed KK
modes disappear on the branes. Here the boundary conditions coming
from the Dirac mass term in the action eq.(\ref{massterm}):
\begin{eqnarray}
\Big(\frac{d}{dy}-m \Big)\,\chi^l_{iLn}(y_0)=0 \nonumber \\
\chi^l_{iRn}(y_0)=0 \, , \label{nLbound}
\end{eqnarray}
where $y_0=0$ or $\pi\,R$. The left handed lepton $\chi^l_{i\,L
n}(y)$ that lives on the visible brane is obtained
\begin{eqnarray}
\chi^l_{iLn}(y)=N_{Ln}\, e^{\sigma/2} \Bigg( J_{\frac{1}{2}-r}
(e^{\sigma}\,x_{nL})+c_L\, Y_{\frac{1}{2}-r}
(e^{\sigma}\,x_{nL})\Bigg)\label{nLmode} \, ,
\end{eqnarray}
by using the Dirac equation for KK mode leptons. Here the constant
$c_L$ is
\begin{eqnarray}
c_L=-\frac{J_{-r-\frac{1}{2}} (x_{nL})}{Y_{-r-\frac{1}{2}}
(x_{nL})} \, .  \label{cL}
\end{eqnarray}
where $N_{Ln}$ is the normalization constant and
$x_{nL}=\frac{m_{Ln}}{k}$. The functions $J_\beta(w)$ and
$Y_\beta(w)$ appearing in eq.(\ref{nLmode}) are the Bessel
function of the first kind and of the second kind, respectively.
The right handed zero mode fields on the branes can be constructed
by considering the $Z_2$ projection condition $Z_2\psi=\gamma_5
\psi$ and this ensures that the right handed zero mode appears,
the right (left) handed KK modes appear (disappear) on the branes
with the boundary conditions:
\begin{eqnarray}
\Big(\frac{d}{dy}+m \Big)\,\chi^E_{iRn}(y_0)=0 \nonumber \\
\chi^E_{iLn}(y_0)=0 \, . \label{nRbound}
\end{eqnarray}
Similarly, the right handed lepton $\chi^E_{i\,R n}(y)$ that lives
on the visible brane is calculated
\begin{eqnarray}
\chi^E_{iRn}(y)=N_{Rn}\, e^{\sigma/2} \Bigg( J_{\frac{1}{2}+r}
(e^{\sigma}\,x_{nR})+c_R\, Y_{\frac{1}{2}+r}
(e^{\sigma}\,x_{nR})\Bigg)\label{nRmode} \, ,
\end{eqnarray}
by using the Dirac equation for KK mode leptons. Here $c_R$ reads
\begin{eqnarray}
c_R=-\frac{J_{r-\frac{1}{2}} (x_{nR})}{Y_{r-\frac{1}{2}} (x_{nR})}
\, , \label{cR}
\end{eqnarray}
where $N_{Rn}$ is the normalization constant and
$x_{nR}=\frac{m_{Rn}}{k}$. Notice that the constant $c_L$, the
$n^{th}$ KK mode mass $m_{Ln}$ in eq.(\ref{nLmode}) and the
constant $c_R$, the $n^{th}$ KK mode mass $m_{Rn}$ in
eq.(\ref{nRmode}) are obtained by using the boundary conditions
eq.(\ref{nLbound}) and eq.(\ref{nRbound}), respectively. For
$m_{L(R)n}\ll k$ and $kR\gg 1$ they are approximated as:
\begin{eqnarray}
m_{Ln}&\simeq&
k\,\pi\,\Big(n-\frac{\frac{1}{2}-r}{2}+\frac{1}{4}\Big)\,
e^{-\pi\,k\,R} \nonumber \, ,\\
m_{Rn} &\simeq&
k\,\pi\,\Big(n-\frac{\frac{1}{2}+r}{2}+\frac{1}{4}\Big)\,e^{-\pi\,k\,R}
\,\,\,\,\,\,\,\, \mbox{for $r<0.5$} \nonumber \, ,\\
m_{Rn}&\simeq&
k\,\pi\,\Big(n+\frac{\frac{1}{2}+r}{2}-\frac{3}{4}\Big)\,e^{-\pi\,k\,R}
\,\,\,\,\,\,\,\, \mbox{for $r>0.5$}\label{mnLR} \, .
\end{eqnarray}

%
\newpage
\newpage
\begin{figure}[htb]
\vskip -3.0truein \centering \epsfxsize=6.8in
\leavevmode\epsffile{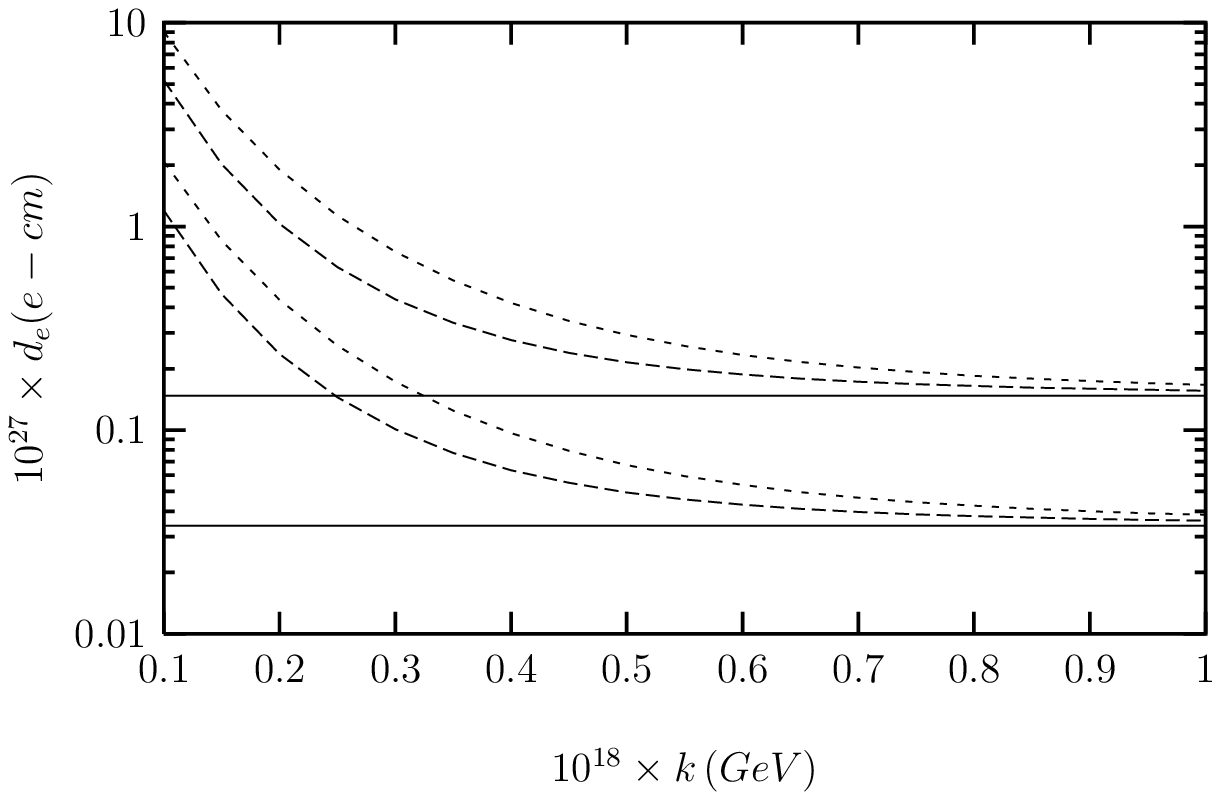} \vskip -3.0truein \caption[]{The
parameter $k$ dependence of the electron EDM $d_e$, for
$\bar{\xi}^{E}_{N,\tau e}=0.01\,(GeV)$. Here, the lower-upper
solid (dashed, small dashed) line represents the $d_e$ for
$sin\,\theta=0.1-0.5$ without KK modes (with KK modes set I, II).}
\label{EDMek}
\end{figure}
\begin{figure}[htb]
\vskip -3.0truein \centering \epsfxsize=6.8in
\leavevmode\epsffile{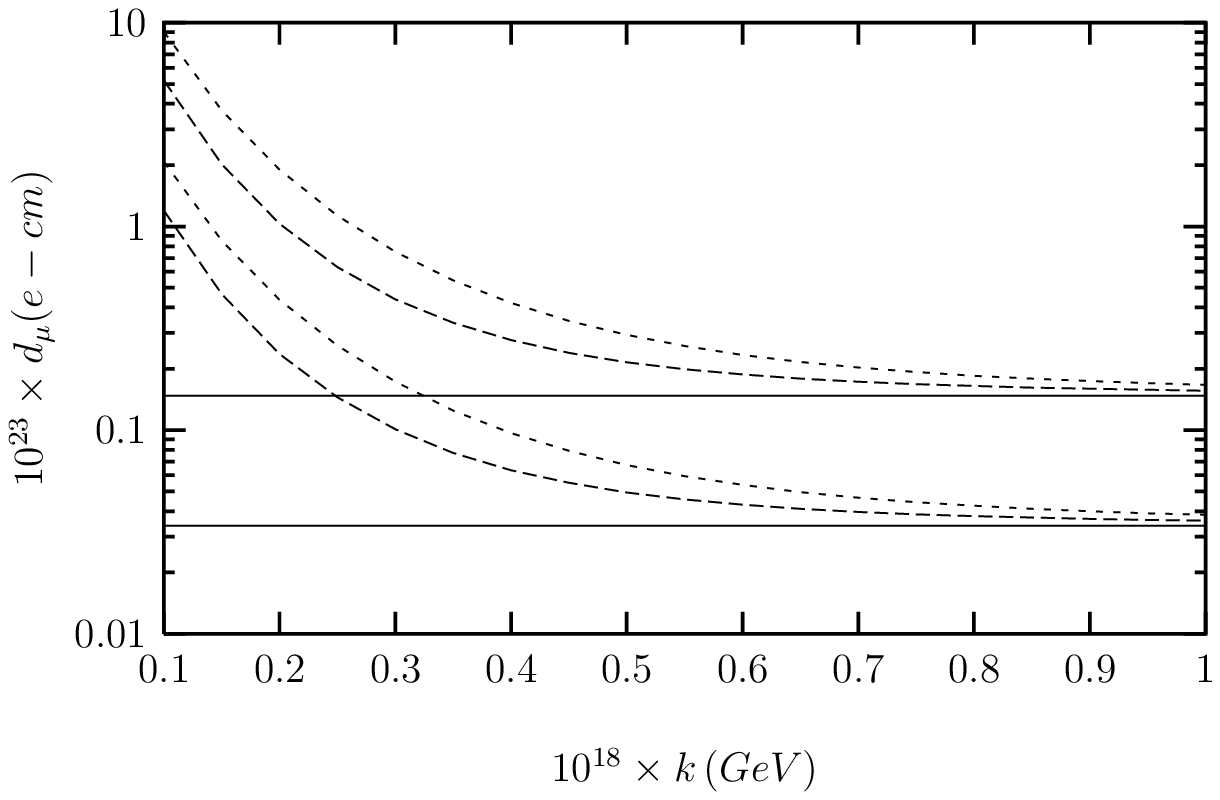} \vskip -3.0truein \caption[]{The
same as Fig. \ref{EDMek} but for $d_\mu$ and
$\bar{\xi}^{E}_{N,\tau \mu}=1\,(GeV)$.} \label{EDMmuk}
\end{figure}
\begin{figure}[htb]
\vskip -3.0truein \centering \epsfxsize=6.8in
\leavevmode\epsffile{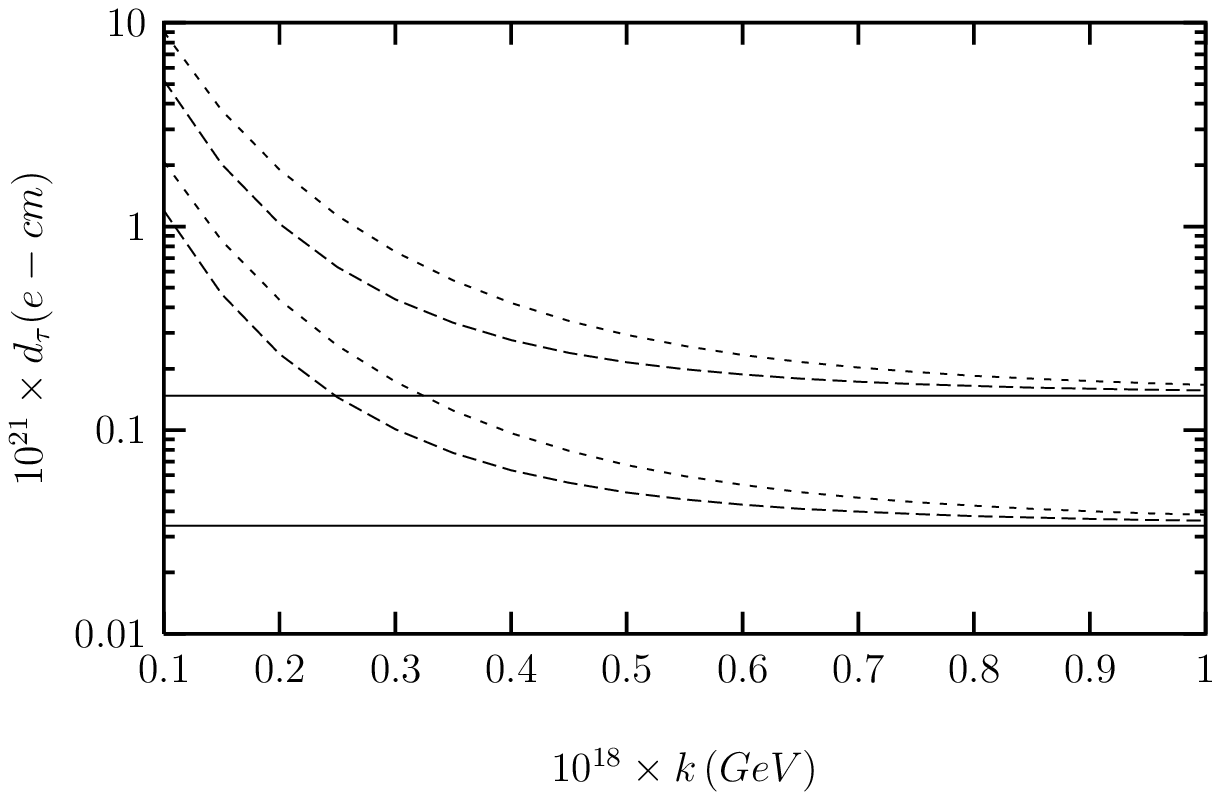} \vskip -3.0truein \caption[]{The
same as Fig. \ref{EDMek} but for $d_\tau$ and
$\bar{\xi}^{E}_{N,\tau \tau}=10\,(GeV)$.} \label{EDMtauk}
\end{figure}
\begin{figure}[htb]
\vskip -3.0truein \centering \epsfxsize=6.8in
\leavevmode\epsffile{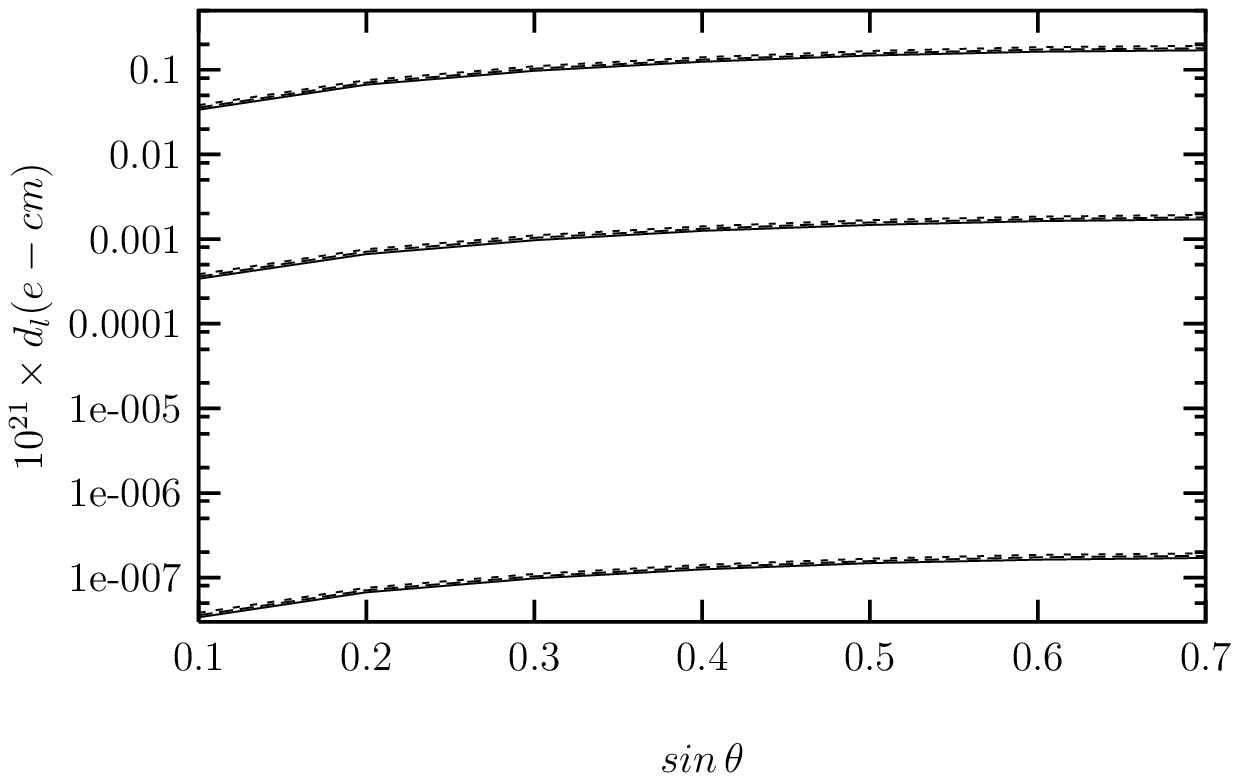} \vskip -3.0truein
\caption[]{The parameter $sin\,\theta$ dependence of the $d_e$;
$d_\mu$; $d_\tau$, for $\bar{\xi}^{E}_{N,\tau e}=0.01\,(GeV)$;
$\bar{\xi}^{E}_{N,\tau \mu}=1.0\,(GeV)$; $\bar{\xi}^{E}_{N,\tau
\tau}=10\,(GeV)$ and for $k=10^{18}\, (GeV)$. Here, the
lower-intermediate-upper solid (dashed, small dashed) line
represents the $d_e$-$d_\mu$-$d_\tau$ without KK modes (with KK
modes set I, II).} \label{EDMemutausintet}
\end{figure}
\end{document}